# Modification of the Kolmogorov-Johnson-Mehl-Avrami rate equation for non-isothermal experiments and its analytical solution

J.Farjas[#], P.Roura

GRMT, Department of Physics, University of Girona, Campus Montilivi, Edif. PII, E17071 Girona, Catalonia, Spain.

**Abstract**

Avrami's model describes the kinetics of phase transformation under the assumption of spatially random nucleation. In this paper we provide a quasi-exact analytical solution of Avrami's model when the transformation takes place under continuous heating. This solution has been obtained with different activation energies for both nucleation and growth rates. The relation obtained is also a solution of the so-called Kolmogorov-Johnson-Mehl-Avrami transformation rate equation. The corresponding non-isothermal Kolmogorov-Johnson-Mehl-Avrami transformation rate equation only differs from the one obtained under isothermal conditions by a constant parameter, which only depends on the ratio between nucleation and growth rate activation energies. Consequently, a minor correction allows us to extend the Kolmogorov-Johnson-Mehl-Avrami transformation rate equation to continuous heating conditions.

# Corresponding author: jordi.farjas@udg.es

## 1. Introduction

Phase transitions are among the most important topics in materials science. Crystallization of amorphous materials and other solid state transformations usually involves nucleation and growth. These transformations are generally described by Kolmogorov-Johnson-Mehl-Avrami [1-5] model. A solution of Kolmogorov-Johnson-Mehl-Avrami's *model* under isothermal conditions can be obtained assuming that nucleation and growth rates are time independent [5]

$$\alpha = 1 - \exp[-(kt)^{m+1}] \quad (1)$$

where $\alpha$ is the transformed phase fraction, $k$ is the overall rate constant that generally depends on temperature, while $m+1$ is usually known as Avrami's exponent ant $t$ is time. For the rest of the paper eqn. (1) will be referred to as the Kolmogorov-Johnson-Mehl-Avrami (KJMA) *relation*. Differentiation of eqn. (1) results in the well known JMA *rate equation*:

$$\frac{d\alpha}{dt} = (m+1)k(1-\alpha) \cdot [-\ln(1-\alpha)]^{m/m+1} \quad (2)$$

Although this equation is obtained from the isothermal solution (eqn. (1)), it constitutes the basis for analyzing non-isothermal experiments [6-8]. This is because the transformation rate "seems" to depend only on temperature (through $k$) and on the transformed fraction. However, depending on the thermal history (e.g. the heating rate), a given value of $\alpha$ will correspond to a different state and consequently it will evolve at a different rate. Indeed, the KJMA rate equation is valid for non-isothermal transformations only when very particular conditions are met (see Section 3.a). Despite these severe limitations, non-isothermal experiments are commonly interpreted within the KJMA rate equation. As pointed out by several authors [7-10], analytical methods based on the KJMA rate equation have been developed regardless of its validity. In particular, the widespread Kissinger method [11] can be applied to any transformation described by the KJMA rate equation [6]. Even though one would expect erroneous conclusions from this incorrect use of the KJMA rate equation, the fact is that good agreement with other exact methods is often obtained. This is a strong indication that many properties of the exact solution are shared by the KJMA rate equation.

In this work a "quasi-exact" solution of Avrami's model for the continuous heating case is obtained by imposing only an Arrhenian temperature dependence for both nucleation and growth rate. Our solution proves to be the exact solution of a KJMA rate equation in which the kinetic constant, *k*, is slightly modified with respect to the isothermal case. So the validity of the KJMA rate equation is extended beyond the severe limitations of the *isothermal* KJMA rate equation.

## 2. The *isothermal* KJMA rate equation

For the transformations involving nucleation and growth and assuming that the nuclei of the new phase are randomly distributed, Avrami [2,3] obtained the following relation

$$\alpha = 1 - \exp[-\alpha_{ex}] \tag{3}$$

where $\alpha_{ex}$ is the extended transformed fraction, i.e. the resulting transformed fraction if nuclei grow through each other and overlap without mutual interference

$$\alpha_{ex}(t) = \int_0^t N(\tau)v(\tau,t)d\tau \tag{4}$$

*N* is the nucleation rate and $v(\tau,t)$ is the volume transformed at time *t* by a single nucleus formed at time $\tau$

$$v(\tau,t) = \sigma\left(\int_\tau^t G(z)dz\right)^m \tag{5}$$

$\sigma$ is a shape factor (e.g. $\sigma = 4\pi/3$ for spherical grains), *G* is the growth rate and *m* depends on the growth mechanism [7,9,12] (e.g. *m=3* for three dimensional growth).

Eqns. (3-5) show the kinetics of the transformation under very general assumptions about the rate constants (any time or temperature dependence) and for any thermal history. The KJMA relation (eqn. (1)) is the particular solution for isothermal conditions provided that both *G* and *N* do not depend on time. The overall rate constant is given by:

$$k = \left(\frac{\sigma N G^m}{m+1}\right)^{1/m+1} \tag{6}$$

In most practical situations it is possible to assume an Arrhenian temperature dependence for both *N* and *G* [12,13]

$$N = N_0 \exp(-E_N/K_B T) \quad \text{and} \quad G = G_0 \exp(-E_G/K_B T) \tag{7}$$

where $E_N$ and $E_G$ are the activation energies for nucleation and growth respectively, and $K_B$ is the Boltzmann constant. Substitution of eqn. (7) into eqn. (6) gives

$$k = \left(\frac{\sigma N_0 G_0^m}{m+1}\right)^{1/m+1} \exp\left(-\frac{E}{K_B T}\right) \equiv k_0 \exp\left(-\frac{E}{K_B T}\right) \tag{8}$$

where we have defined the overall activation energy as:

$$E \equiv \frac{E_N + mE_G}{m+1} \quad (9)$$

As mentioned in the introduction, differentiation of eqn. (1) leads to eqn. (2) which will be referred to as the *isothermal KJMA rate equation* for the rest of the paper.

## 3. The non-isothermal case

Although non-isothermal experiments can use any arbitrary thermal history, the most usual experiments performed in thermal analysis involve heating at a constant rate, $\beta = dT/dt$. Therefore, and for the rest of the paper, we will deal with this particular non-isothermal condition.

*3.a The isokinetic case ($E_N=E_G$).*

The KJMA rate equation can be applied to the non-isothermal case when the transformation rate depends exclusively on temperature and on the degree of transformation [6,8,9] and not on the thermal history. This condition is fulfilled in particular cases such as "site saturation", where nucleation is completed prior to crystal growth [9,14], or the singular "isokinetic" situation where $N$ and $G$ have the same activation energy [2].

For a constant heating rate, introducing eqn. (7) into eqn. (5) gives the volume transformed by a single nucleus:

$$v(\tau,t) = \sigma \left(\frac{G_0 E_G}{\beta K_B}\right)^m \left[ p\left(\frac{E_G}{K_B T}\right) - p\left(\frac{E_G}{K_B T'}\right) \right]^m \quad (10)$$

where $T=T_0+\beta t$, $T'=T_0+\beta\tau$, ($T_0$ is the initial temperature) and the function $p(x)$ is defined as (see Appendix A):

$$p(x) \equiv \int_x^\infty \frac{\exp(-u)}{u^2} du \quad (11)$$

Accordingly, the extended transformation fraction, $\alpha_{ex}$, can be deduced after substituting eqn. (10) into eqn. (4), and assuming that the transformation rate is negligible at $T_0$:

$$\alpha_{ex}(t) = \sigma G_0^m N_0 \left(\frac{E_G}{\beta K_B}\right)^{m+1} \int_x^\infty \exp\left(-\frac{E_N}{E_G}u\right)\frac{1}{u^2}[p(x)-p(u)]^m du \qquad (12)$$

where $x \equiv E_G/K_B T$. This integral can be solved analytically when $E_N=E_G$ (isokinetic case). Indeed, by substituting eqn. (A.3) into eqn. (12) one obtains

$$\alpha_{ex}(t) = \sigma G_0^m N_0 \left(\frac{E}{\beta K_B}\right)^{m+1} \int_x^\infty \left(-\frac{dp(u)}{du}\right)[p(x)-p(u)]^m du = \left[k_0 \frac{E}{\beta K_B} \cdot p\left(\frac{E}{K_B T}\right)\right]^{m+1} \qquad (13)$$

and, substituting $\alpha_{ex}$ into eqn. (3) gives, finally, the transformed fraction:

$$\alpha = 1 - Exp\left\{-\left[k_0 \frac{E}{\beta K_B} p\left(\frac{E}{K_B T}\right)\right]^{m+1}\right\} \qquad (14)$$

Derivation of eqn. (14) with respect to time shows that it is an exact solution of the isothermal KJMA rate equation. Consequently, this equation is valid for non-isothermal conditions provided that $E_N=E_G$ (isokinetic case). The literature shows [10,15] that the solution, $\alpha(t)$, for the "site saturation" case also obeys the isothermal KJMA rate equation (eqn. (2)).

*3.b The general case ($E_N \neq E_G$).*

When $E_N \neq E_G$, the integral of eqn. (12) has no analytical solution. For most experiments $E/K_B T >> 1$, thus $p(x)$ is usually approximated by its first term in a series of $1/x$ [7,10,16-21] (see Appendix A):

$$p(x) \approx \frac{\exp(-x)}{x^2} \qquad (15)$$

with this first order approximation, a number of authors [18-21] obtained an identical solution of eqn. (12).

We will follow a different approach to solve the integral of eqn. (12). The fact that the arguments of the exponential functions and $p(x)$ are different makes it impossible to solve it analytically. We overcome this problem by replacing these arguments by a common averaged argument (see Appendix B). With this approximation, eqn. (12) can be solved and the corresponding transformed fraction is (see Appendix C):

$$\alpha = 1 - Exp\left\{-\left[k_0 C \frac{E}{\beta K_B} p\left(\frac{E}{K_B T}\right)\right]^{m+1}\right\} \qquad (16)$$

where $C$ is a constant that depends on $m$, $E_N$ and $E_G$:

$$C \equiv \left( \frac{(m+1)! E^{m+1}}{\prod_{i=0}^{m}(E_N + iE_G)} \right)^{1/(m+1)} \tag{17}$$

It is worth noting that our approximate solution coincides with the exact isokinetic solution (eqn. (14)), except for the constant $C$. As a consequence, the approximate solution for the general non-isothermal case is also a solution of the isothermal KJMA rate equation with the overall rate constant $k$ multiplied by the constant $C$:

$$\frac{d\alpha}{dt} = (m+1) \cdot C \cdot k \cdot (1-\alpha) \cdot [-\ln(1-\alpha)]^{m/(m+1)} \tag{18}$$

In the rest of the paper, we will refer to eqn. (18) as the *non-isothermal KJMA rate equation*. As expected when $E_N = E_G$, $C$ reduces to unity and our solution (eqn. (16)) coincides with the exact solution for this particular limit (eqn. (14)).

*3.c Accuracy of the non-isothermal KJMA rate equation*

We will analyze the accuracy of our solution by comparing it to the exact solution that results from the numerical integration of Avrami's model (eqn. (3-5)). We will also show that it is much more accurate than: (a) the solution of the isothermal KJMA rate equation (eqn. (2)) with $E = (E_N + mE_G)/(m+1)$ and; (b) that of Vazquez et al. [18] and other authors [19-21] (eqn. D.1), which consists in replacing $p(x)$ by its first order approximation $\exp(-x)/x^2$ in eqn. (16) (see Appendix D). In fact, one can easily state that the relative error in calculating $\alpha_{ext}$ with respect to the exact solution for the aforementioned solutions only depends on two parameters: the ratio $E_N/E_G$ and the value of $E/K_BT$. When this error is small (say <0.2), the same conclusion applies to the relative error of $\alpha$. A linear expansion of eqn. (3) provides the relation between the relative error of $\alpha$ ($\Delta\alpha/\alpha$) and that of $\alpha_{ex}$ ($\Delta\alpha_{ex}/\alpha_{ex}$):

$$\left( \frac{\Delta\alpha}{\alpha} \right) = \frac{(1-\alpha) \cdot [-\ln(1-\alpha)]}{\alpha} \left( \frac{\Delta\alpha_{ex}}{\alpha_{ex}} \right) \tag{19}$$

In Fig. 1, $\Delta\alpha/\alpha$ has been plotted versus $E_N/E_G$ for different "normalized" temperatures ($K_BT/E$). From Fig. 1 it can be concluded that in any condition our solution is the most accurate one. For $\alpha=0.5$, Fig. 1 tells us that the relative error of our solution is lower than 0.07 for $E/K_BT$ greater than 20. The absolute error $\Delta\alpha$ is thus lower than 0.035. In most practical situations $E/K_BT$ is greater than 30, thus $\Delta\alpha$ would be even lower (<0.01). We conclude that with the accuracy of experimental data, our solution can barely be distinguished from the exact solution. It is worth mentioning that Fig. 1 fails to give the correct result for the isokinetic case ($E_N/E_G=1$), even though the approximation of Vázquez et al. (curve b) is more accurate than the solution of the isothermal KJMA rate equation (curve a).

In the limit where $E_G >> E_N$, nucleation takes place before growth [13] and the "site saturation" approximation is the most appropriate description. Conversely, when $E_N >> E_G$ either heterogeneous nucleation dominates and the "site saturation" approximation is also the appropriate description, or crystallization is driven by epitaxial growth.

Another way to test the accuracy is by plotting the crystallization rate as a function of temperature. This has been done for the particular $G$ and $N$ values of amorphous silicon crystallization [22]. The result has been plotted in Fig. 2. It is clear that our solution and the exact solution are practically indistinguishable (the shift of the peak temperature is less than 0.1 ºC). Moreover, although the isothermal KJMA rate equation is not an accurate solution, it predicts the correct peak shape since the only difference with respect to the non-isothermal KJMA rate equation is a constant factor.

## 4. Kinetic analysis of phase transformations under conditions of a constant heating rate

Most of the analyses of thermoanalytical experiments are based on the isothermal KJMA equation [6-8,17] (eqn. (2)). Since our solution obeys a formally identical equation (eqn. (18)) for the non-isothermal case, this means that these analyses can also be applied to non-isothermal experiments provided that the overall rate constant $k$ of the isothermal KJMA rate equation is modified by the constant factor $C$.

A common feature of most thermoanalytical techniques is the identification of two parameters which produce a straight line when plotted against each other. The activation energy $E$ and the transformation order are obtained from this representation. In general, most methods are isoconversional, i.e. these two parameters are related to determining the temperatures at which an equivalent state of conversion is reached.

The most widespread method is the Kissinger method [11] where $\ln(\beta/T_P^2)$ is plotted versus $1/T_P^2$ ($T_P$ is the peak temperature, i.e. $\left.\frac{d^2\alpha}{dt^2}\right|_{T_P} = 0$). The Kissinger plot relies on the linear relationship;

$$\ln\frac{\beta}{T_P^2} = \frac{A}{T_P} + B \tag{20}$$

The activation energy can be obtained from $A$, whereas $B$ contains information about the pre-exponential factor of the kinetic constant. For the non-isothermal KJMA rate equation it can be shown that (Appendix F):

$$A = -\frac{E}{K_B} \quad \text{and} \quad B = \ln\left[-\frac{K_B}{E}k_0 C g'(\alpha_P)\right] \tag{21}$$

where $g'(\alpha)$ is:

$$g'(\alpha) = m \cdot [-\ln(1-\alpha)]^{\frac{-1}{m+1}} - (m+1)\cdot[-\ln(1-\alpha)]^{\frac{m}{m+1}} \tag{22}$$

and $\alpha_P = \alpha(T_P)$. The value of $\alpha_P$ can be obtained for the non-isothermal KJMA rate equation (eqn. (18)) after calculating the second derivative of its exact solution (eqn. (16)) and equating it to zero:

$$-(m+1)(k_0 C)^{m+1}\left(\frac{E}{\beta K_B}\right)^m \left[p\left(\frac{E}{K_B T_P}\right)\right]^{m+1} e^{\left(-\frac{E}{K_B T_P}\right)} + m\frac{\beta K_B}{E}e^{\left(-\frac{E}{K_B T_P}\right)} + \beta\frac{E}{K_B T_P^2}p\left(\frac{E}{K_B T_P}\right) = 0$$

by substituting $p(x)$ by its first order approximation in the last term, one gets:

$$\left[k_0 C \left(\frac{E}{\beta K_B}\right) p\left(\frac{E}{K_B T_P}\right)\right]^{m+1} \approx 1 \tag{23}$$

which tells us that at the peak temperature, $\alpha_{ex}=1$, (compare eqn. (23) with eqn. (16)) and consequently:

$$\alpha_P \approx 1 - e^{-1} = 0.632 \tag{24}$$

The previous result does not depend on any parameter and it was proposed by Henderson [6] as a test for the applicability of the KJMA model. Moreover, $g'(\alpha_P) = -1$, so the Kissinger constant term is reduced to $B = \ln\left[\frac{K_B}{E}k_0 C\right]$.

Deviations of $\alpha_P$ from 0.632 are negligible in real situations ($E/K_B T_P > 20$) [8]. The prediction of the peak temperature for the non-isothermal KJMA rate equation is plotted in Fig. 3 and compared with the exact solution. For a wide range of heating rates (1<$\beta$<100 K/min), discrepancies vary from 0.11 to 0.14 K. Consequently, within experimental accuracy, eqn. (18) can be considered exact.

The Coats-Matusita method [23,24] can be worked out from eqn. (16) by taking twice the logarithm:

$$\ln[-\ln(1-\alpha)] = (m+1)\ln\left(p\left(\frac{E}{K_B T_P}\right)\right) + (m+1)\ln\left(k_0 C \frac{E}{\beta K_B}\right) \qquad (25)$$

By substituting $p(x)$ by its first order approximation, we obtain

$$\ln[-\ln(1-\alpha)] = (m+1)\ln\left(e^{-\frac{E}{K_B T}} \Big/ \left(\frac{E}{K_B T}\right)^2\right) + (m+1)\ln\left(k_0 C \frac{E}{\beta K_B}\right) \qquad (26)$$

It can be easily verified numerically that the plot of $\ln(p(y))$ versus $y$ for 20<$y$<60 exhibits a clear linear trend: $\ln(p(y)) \approx -5.2813 - 1.051y$ [17]. Therefore, eqn. (26) can be reduced to:

$$\ln[-\ln(1-\alpha)] = -1.051 \cdot (m+1)\frac{E}{K_B T} + (m+1)\ln\left(k_0 C \frac{E}{\beta K_B}\right) - 5.2813 \qquad (27)$$

Thus, the plot of $\ln[-\ln(1-\alpha)]$ as a function of reciprocal temperature is linear with a slope of $-1.051 \cdot (m+1)E/K_B$. Following a similar procedure, we can verify the validity of the method developed by Piloyan [25] in which the activation energy $E$ is obtained from the slope of the linear relationship between $\ln(d\alpha/dt)$ and $1/T$.

Most of the abovementioned isoconversional methods rely on replacing $p(x)$ at a given stage by the first term of its series (eqn. (15)). However, more accurate isoconversional methods do not use any mathematical approximation [26] or are based on a more precise approximation [8,17,27,28]. Nonetheless, since our solution is an exact solution of the KJMA rate equation, their applicability is automatically extended to the general non-isothermal case.

Ozawa's method [29] is a widely used exact method, and can be inferred by taking twice the logarithm of eqn. (16)

$$Log_{10}[-\ln(1-\alpha)] = (m+1)Log_{10}\left[k_0 C \frac{E}{K_B} p\left(\frac{E}{K_B T}\right)\right] - (m+1)Log_{10}(\beta) \qquad (28)$$

Thus, the slope of the plot $Log_{10}[-\ln(1-\alpha)]$ against $Log_{10}(\beta)$ at a given temperature yields the value of *m*.

## 5. Conclusion

We have obtained the more accurate approximate solution of Avrami's model under conditions of constant temperature scan rate. This solution is also a solution of the KJMA rate equation. Indeed, it is the solution of an equation that differs from the *isothermal KJMA rate equation* by a constant factor. As a consequence, most of the classical kinetic analysis techniques, which are based on the KJMA rate equation, are valid for the general non-isothermal case. In view of this result, one can understand the noteworthy success and satisfactory results obtained from methods based on KJMA regardless of the fact that they rely on the *incorrect* assumption of the validity of the isothermal KJMA equation for the non-isothermal case.

Within experimental accuracy, the *non-isothermal KJMA rate equation* and its analytical solution can be considered as a quasi-exact description of transformation kinetics. Table I summarizes the main results obtained in this paper.

**Appendix A: the function *p(x)***

The function

$$p(x) \equiv \int_x^\infty \frac{\exp(-u)}{u^2} du \qquad (A.1)$$

is related to the exponential integral $E_2(x)$ [30] according to:

$$p(x) = \frac{E_2(x)}{x} \qquad (A.2)$$

Its first derivative is

$$\frac{d}{dx} p(x) = -\frac{\exp(-x)}{x^2} \qquad (A.3)$$

and, from the asymptotic expansion of $E_2(x)$ [30], $p(x)$ can be developed as

$$p(x) \sim \frac{\exp(-x)}{x^2} \sum_{i=0} (-1)^i \frac{(i+1)!}{x^i} \qquad (A.4)$$

**Appendix B: approximate solution for** $\int_x^\infty [p(bu)]^n \frac{\exp(-u)}{u^2} du$

In this appendix we will state that relation B.1 is exact for the first order asymptotic expansion of $p(x)$,

$$I = \int_x^\infty [p(bu)]^n \frac{\exp(-u)}{u^2} du \approx \left(\frac{nb+1}{n+1}\right)^{2n+2} \frac{1}{b^{2n}(nb+1)} \left[p\left(\frac{nb+1}{n+1}x\right)\right]^{n+1} \qquad (B.1)$$

With the new variable $t=(nb+1)\,u\,/(n+1)$, the integral $I$ can be written as

$$I = \frac{nb+1}{n+1} \int_{\frac{nb+1}{n+1}x}^\infty \left[p\left(\frac{nb+b}{nb+1}t\right)\right]^n \frac{1}{t^2} \exp\left(-\frac{n+1}{nb+1}t\right) dt \qquad (B.2)$$

Next, we perform the following approximation which is exact for the first term in the series expansion of $p(x)$ (eqn. (A.4))

$$p\left(\frac{nb+b}{nb+1}t\right) \approx \left(\frac{nb+1}{nb+b}\right)^2 p(t) \exp\left(\frac{1-b}{nb+1}t\right) \qquad (B.3)$$

then the integral $I$ is approximately

$$I \approx \left(\frac{nb+1}{n+1}\right)^{2n+1} \frac{1}{b^{2n}} \int_{\frac{nb+1}{n+1}x}^\infty [p(t)]^n \frac{\exp(-t)}{t^2} dt \qquad (B.4)$$

bearing in mind relation A.3, the latter integral can be solved and is the right hand side term of B.1.

**Appendix C: Approximate non-isothermal solution of Kolmogorov-Johnson-Mehl-Avrami's model**

If we integrate by parts, eqn. (12) becomes:

$$\alpha_{ex}(t) = \sigma G_0^m N_0 \left(\frac{E_G}{\beta K_B}\right)^{m+1} m \int_x^\infty \frac{\exp(-u)}{u^2} [p(x) - p(u)]^{m-1} \left(\int_u^\infty \frac{1}{z^2} \exp(-az) dz\right) du = $$
$$= \sigma G_0^m N_0 \left(\frac{E_G}{\beta K_B}\right)^{m+1} ma \int_x^\infty \frac{\exp(-u)}{u^2} [p(x) - p(u)]^{m-1} p(au) du \quad (C.1)$$

Where $a \equiv E_N/E_G$. After a second integration by parts, $\alpha_{ex}$ is expressed as:

$$\alpha_{ex}(t) = \sigma G_0^m N_0 \left(\frac{E_G}{\beta K_B}\right)^{m+1} am(m-1) \int_x^\infty \frac{\exp(-u)}{u^2} [p(x) - p(u)]^{m-2} \left(\int_u^\infty \frac{\exp(-z)}{z^2} p(az) dz\right) du$$

By substituting the approximation (B.1) one obtains

$$\alpha_{ex}(t) \approx \sigma G_0^m N_0 \left(\frac{E_G}{\beta K_B}\right)^{m+1} m(m-1) \left(\frac{a+1}{2}\right)^4 \frac{1}{a(a+1)} \int_x^\infty \frac{\exp(-u)}{u^2} [p(x) - p(u)]^{m-2} \left[p\left(\frac{a+1}{2}u\right)\right]^2 du$$

After $m$ times of first integration by parts and if we substitute (B.1) again, $\alpha_{ex}$ reduces to:

$$\alpha_{ex}(t) \approx \sigma G_0^m N_0 \left(\frac{E_G}{\beta K_B}\right)^{m+1} \frac{m!}{\prod_{i=0}^m (a+i)} \left(\frac{a+m}{m+1}\right)^{2m+2} \left[p\left(\frac{a+m}{m+1} x\right)\right]^{m+1}$$

If $a$ is replaced by $E_N/E_G$ the later expression can be rewritten as:

$$\alpha_{ex}(t) = \left[k_0 C \frac{E}{\beta K_B} p\left(\frac{E}{K_B T}\right)\right]^{m+1} \quad (C.2)$$

where $C \equiv \left(\frac{(m+1)! E^{m+1}}{\prod_{i=0}^m (E_N + iE_G)}\right)^{1/(m+1)}$.

**Appendix D. Approximate solution of Vázquez et al.**

The solution of Vázquez et al. [18]:

$$\alpha_{ex}(t) = \sigma K_B \left(\frac{K_B}{E_G}\right)^m \left(\sum_{i=0}^{m}(-1)^i \binom{m}{i}\frac{1}{E_N + iE_G}\right)\left[\frac{(G_0^m N_0)^{1/m+1} T^2}{\beta}\exp\left(-\frac{E}{K_B T}\right)\right]^{m+1} \quad (D.1)$$

can be deduced from our solution (eqn. (16)) simply by substituting *p(x)* by its first order approximation:

$$p\left(\frac{E}{K_B T}\right) \approx \frac{\exp\left(-E/K_B T\right)}{\left(E/K_B T\right)^2}$$

and rewriting *C* as (see Appendix E):

$$C = \left((m+1)\frac{E^{m+1}}{E_G^m}\sum_{i=0}^{m}(-1)^i \binom{m}{i}\frac{1}{E_N + iE_G}\right)^{1/m+1}$$

One can easily come to the same conclusion for the solutions obtained by other authors [19-21].

**Appendix E. Proof of the identity** $\dfrac{n!}{\prod_{i=0}^{n}(x+i)} = \sum_{i=0}^{n}(-1)^i \binom{n}{i}\dfrac{1}{x+i}$

The left hand side term can be developed in simple fractions

$$\frac{n!}{\prod_{i=0}^{n}(x+i)} = n!\sum_{i=0}^{n}\frac{a_i}{x+i} \quad (D.1)$$

The identity is then established if we can prove that:

$$a_i = \frac{1}{n!}(-1)^i \binom{n}{i} = (-1)^i \frac{1}{i!(n-i)!} \quad (D.2)$$

Expansion of the right hand side term of (D.1) results in

$$\frac{n!}{\prod_{i=0}^{n}(x+i)} = n!\frac{\sum_{i=0}^{n}a_i \prod_{j\neq i}(x+j)}{\prod_{i=0}^{n}(x+i)} \quad (D.3)$$

which implies

$$1 = \sum_{i=0}^{n}a_i \prod_{j\neq i}(x+j) \quad (D.4)$$

Let's evaluate the previous relation for *x = -i, i = 0, ..., n*. For each of these values all the terms of the sum are equal to zero except one, the only one that contains the factor *x+i*, i.e.:

$$1 = a_i \prod_{j \neq i}(-i+j) = a_i(-i)\cdot(-i+1)\cdots(-1)\cdot 1\cdot 2\cdots(-i+n) = a_i(-1)^i i!(n-i)!$$

Thus relation D.2 and consequently the identity are proved.

**Appendix F. Kissinger plot.**

Consider a rate equation with the general form:

$$\frac{d\alpha}{dt} = K \cdot g(\alpha) \quad \text{with} \quad K = K_0 e^{-\frac{E}{K_B T}} \quad \text{(F.1)}$$

where $g$ is an arbitrary function. For a constant heating rate experiment $(T = T_0 + \beta t)$, the peak temperature $T_P$ is determined by the condition:

$$\left.\frac{d^2\alpha}{dt^2}\right|_{T_P} = 0 \quad \text{(F.2)}$$

which leads to the relationship:

$$\frac{\beta}{K_B T_P^2}\frac{1}{g'(\alpha_P)} + \frac{K_0}{E}e^{-\frac{E}{K_B T_P}} = 0 \quad \text{(F.3)}$$

where $g'(\alpha_P)$ is the first derivative of $g$ with respect to $\alpha$ evaluated at the maximum of $d\alpha/dt$. Equation (F.3) can be easily transformed into a form that is suitable for a Kissinger plot:

$$\ln\frac{\beta}{T_P^2} = -\frac{E}{K_B T_P} + \ln\left[-\frac{K_B}{E}K_0 g'(\alpha_P)\right] \quad \text{(F.4)}$$

**Acknowledgements**

We whish to thank Dr. Francesc-Xavier Massaneda for developing appendix E. This work was supported by the Spanish Programa Nacional de Materiales under contract MAT-2002-04236-C04-02.

**Table 1**. Summary of the main results

| | |
|---|---|
| Non-isothermal KJMA rate equation | $\dfrac{d\alpha}{dt} = (m+1) \cdot C \cdot k \cdot (1-\alpha) \cdot [-\ln(1-\alpha)]^{m/(m+1)}$ |
| Its solution | $\alpha = 1 - Exp\left\{-\left[k_0 C \dfrac{E}{\beta K_B} p\left(\dfrac{E}{K_B T}\right)\right]^{m+1}\right\}$ |
| Time independent rate constant | $k_0 = \left(\dfrac{\sigma N_0 G_0^m}{m+1}\right)^{1/(m+1)}$ |
| Correction factor | $C \equiv \left(\dfrac{(m+1)!\, E^{m+1}}{\prod\limits_{i=0}^{m}(E_N + i E_G)}\right)^{1/(m+1)}$ |
| Activation energy | $E = \dfrac{E_N + m E_G}{m+1}$ |
| Kissinger constant term | $B = \ln\left[\dfrac{K_B}{E} k_0 C\right]$ |

**Figure captions**

Figure 1. Solid curves: relative error in calculating the transformed fraction from our solution (non-isothermal KJMA equation, eqn. (16)) with respect to the exact solution (eqn. (12)) for three different values of $E/K_BT$. Dashed curves: relative error in calculating the transformed fraction using (a) the isothermal KJMA rate solution (eqn. (2)) and (b) the first order solution of Vazquez et al. [18] (eqn. (D.1)) for $E/K_BT=20$. All curves have been calculated for a transformed fraction of 0.5.

Figure 2. Three–dimensional crystallization rate of amorphous silicon as a function of temperature calculated from: the exact solution (solid curve), our solution eqn. (18) (dotted curve), the isothermal KJMA rate equation (curve (a)) and from eqn. D.1 (curve (b)) for a heating rate of 40 K/min. Experimental parameters [22]: $E_G=3.1$ eV, $E_N=5.3$ eV, $G_0=1.6\ 10^7$ m/s, $N_0=1.5\ 10^{44}$ s$^{-1}$m$^{-3}$, $m=3$ and $\beta=40$ K/min.

Figure 3. Kissinger plot corresponding to the crystallization of amorphous silicon calculated from: the exact solution (solid curve), our solution (dotted curve), the isothermal KJMA rate equation (curve (a)) and from eqn. D.1 (curve (b)) for a heating rate ranging form 1 to 100 K/min. Experimental parameters [22]: $E_G=3.1$ eV, $E_N=5.3$ eV, $G_0=1.6\ 10^7$ m/s, $N_0=1.5\ 10^{44}$ s$^{-1}$m$^{-3}$ and $m=3$.

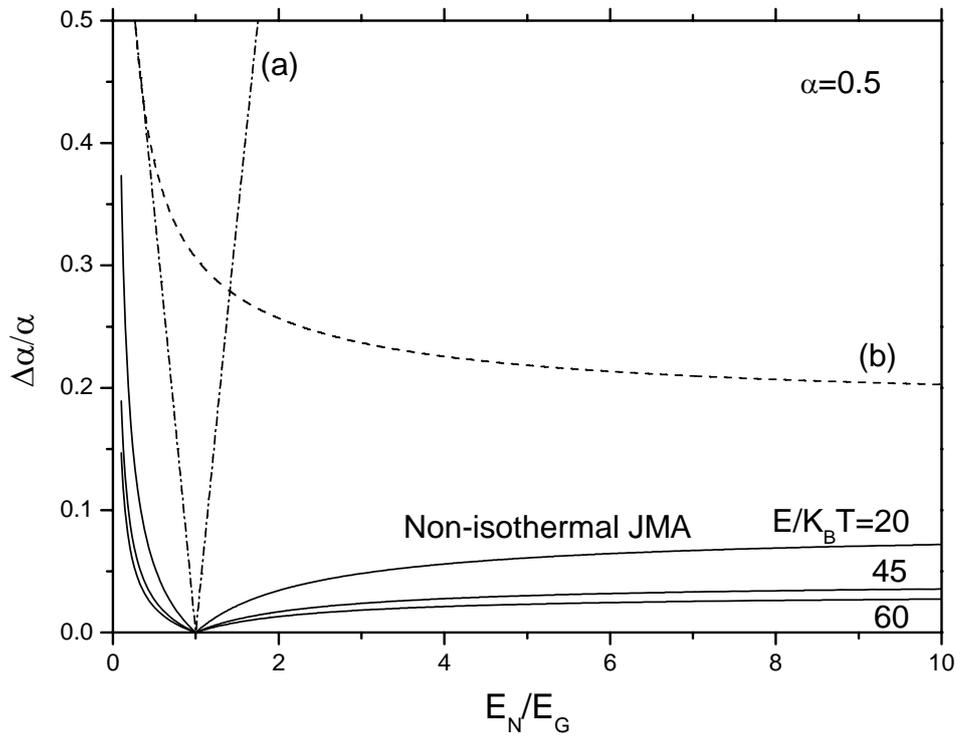

Fig. 1

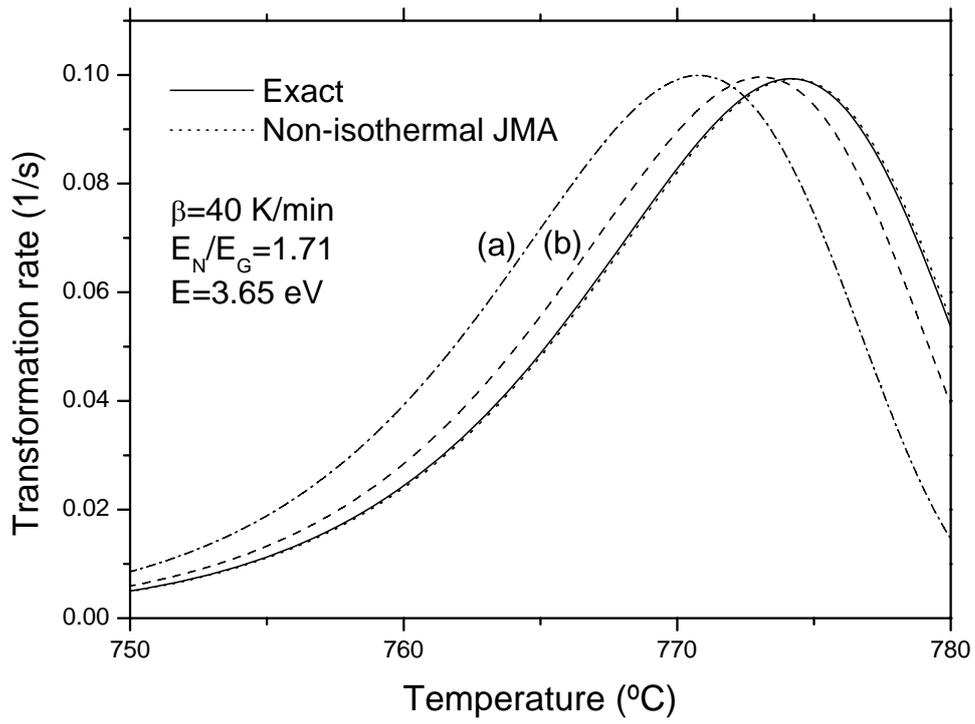

Fig. 2.

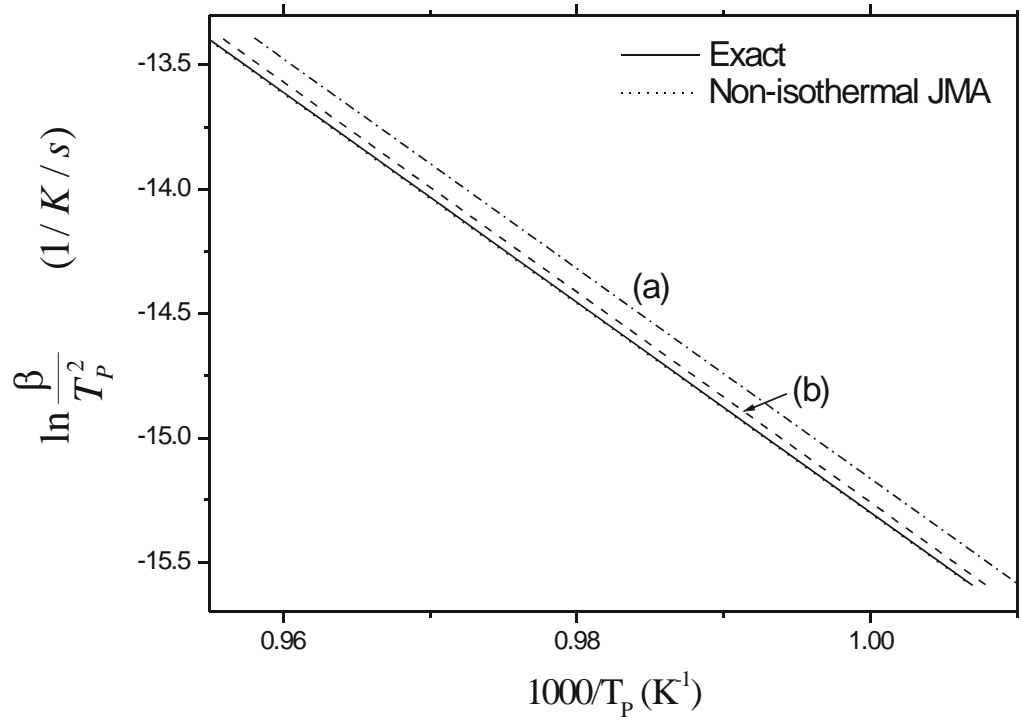

Fig. 3.